\documentclass[usenatbib]{mn2e}
\usepackage{epsfig}
\usepackage{amsmath}
\usepackage{hyperref}
\hypersetup{
 colorlinks,
citecolor = [rgb]{0,0.5,0} }
\usepackage{amsmath}
\def\gtwid{\mathrel{\raise.3ex\hbox{$>$\kern-.75em\lower1ex\hbox{$\sim$}}}}
\def\ltwid{\mathrel{\raise.3ex\hbox{$<$\kern-.75em\lower1ex\hbox{$\sim$}}}}
\def\\{\hfil\break}

\def\eg{{\it e.g.}}
\def\etal{{\it et al.}}
\newcommand{\degree}{{\rm o}}

\def\lesssim{\mathrel{\hbox{\rlap{\hbox{\lower2pt\hbox{$\sim$}}}\raise2pt\hbox{$<$}}}}
\def\gtrsim{\mathrel{\hbox{\rlap{\hbox{\lower2pt\hbox{$\sim$}}}\raise2pt\hbox{$>$}}}}

\newcommand{\mamo}[1]{\mbox{$#1$}}
\newcommand{\unit}[1]{\ifmmode \:\mbox{\rm #1}\else \mbox{#1}\fi}
\newcommand{\mone}{\mamo{^{-1}}}

\newcommand{\mpc}{\unit{Mpc}}

\newcommand{\hmpc}{\mamo{h\mone}\mpc}

\begin{document}

\title[Bulk Flows from the CF4]{Analyzing the Large-Scale Bulk Flow using CosmicFlows4: Increasing Tension with the Standard Cosmological Model}
\vskip 0.5cm
\author[Watkins \etal]{Richard Watkins$^{\dagger,1}$, Trey Allen$^{\dagger}$, Collin James Bradford$^{\dagger}$, Albert Ramon Jr.$^{\dagger}$, \and Alexandra Walker$^{\dagger}$, Hume A. Feldman$^{\star,2}$, Rachel 
Cionitti$^\star$,  Yara Al-Shorman$^\star$, \and Ehsan Kourkchi$^{\dagger\dagger}$, \& R. Brent Tully$^{\dagger\dagger}$\\
$^\dagger$Department of Physics, Willamette University, Salem, OR 97301, USA.\\
$^\star$Department of Physics \& Astronomy, University of Kansas, Lawrence, KS 66045, USA.\\
$^{\dagger\dagger}$Institute for Astronomy, University of Hawaii, Honolulu, HI 96822, USA.\\
emails: $^1$rwatkins@willamette.edu;\, $^2$feldman@ku.edu}

\maketitle
\begin{abstract}
We present an estimate of the bulk flow in a volume of radii $150-200h^{-1}$Mpc using the minimum variance (MV) method with data from the \textit{CosmicFlows-4} (CF4) catalog.   The addition of new data in the CF4 has resulted in an increase in the estimate of the bulk flow in a sphere of radius $150h^{-1}$Mpc relative to the \textit{CosmicFlows-3} (CF3). This bulk flow has less than a $0.03\%$ chance of occurring in the Standard Cosmological Model ($\Lambda$CDM) with cosmic microwave background derived parameters. Given that the CF4 is deeper than the CF3, we were able to use the CF4 to accurately estimate the bulk flow on scales of $200h^{-1}$Mpc (equivalent to 266 Mpc for Hubble constant $H_o=75$ km/s/Mpc) for the first time. This bulk flow is in even greater tension with the Standard Model, having less than $0.003\%$ probability of occurring. To estimate the bulk flow accurately, we introduce a novel method to calculate distances and velocities from distance moduli that is unbiased and accurate at all distances.  Our results are completely independent of the value of $H_o$.  
\end{abstract}

\section{Introduction}
\label{sec:intro}


In an expanding universe, the observed redshift of an object at cosmological distances arises from two separate and independent effects. The first  is due to the expansion of the universe and is proportional to the distance to the object \citep{Hub1929}. The other is due to the local (peculiar) velocity that is determined only by the mass distribution around the object. Peculiar velocities \citep{Zwicky1933,RubFor70,ZelSun1980} have been used as a probe of large-scale structure, as they provide information about density perturbations, and hence the mass distribution, on scales of and larger than the effective depths of surveys. To study the mass distribution, one must take into account redshift distortions \citep[RSD][]{Ham1998} that arise from peculiar velocities and thus bias results from redshift surveys. Furthermore, peculiar velocity studies provide a mechanism to identify the possible sources of gravitation attraction in large volumes \citep[\eg][]{JacBraCiarDav92,Wil94,StrWil95}. Many groups have surveyed \citep[\eg][]{RubFor70,RubThoForRob76,DreFabBurDav87,LP94,RiePreKir95,ZarBerdaC01,KasAtrKoc08,SprMagCol14,TulCouDol13,CF3} and analyzed \citep[\eg][]{FelWat94,WatFel95,NusDav95,FelWat98,FelWatMelCha03,HuiGree06,SarFelWat07,WatFel07,FelWat08,Tul08,AbaErd09,WatFelHud09,FelWatHud10,RatKovItz11,DavNusMas11,NusDav11,NusBraDav11,AgaFelWat12,MacFelFerJaf12,TurHudFel12,Nusser14,CarTurLav15,WatFel15a,WatFel15,Nusser16,HelNusFeiBil17,PeeWatFel18,WanRooFelFel2018}
peculiar velocity catalogs in the last half a century. 

To construct a peculiar velocity catalog, one must estimate both the redshift of an object and its distance, usually in the form of the distance modulus. The redshift to galaxies is a reasonably straightforward and accurate measurement, usually using spectroscopic data \citep[\eg][]{GelHuc1989}. However, estimating the distance modulus to faraway galaxies is difficult, expensive, and requires a great deal of telescope time.   The most common distance indicators, the Tully-Fisher \citep[TF,][]{TF77} and the Fundamental Plane \citep[FP,][]{dressler87} relations, have large uncertainties,  around 0.4 in the distance modulus for individual galaxies.  Other, more accurate, distance indicators exist,  such as SNIa \citep[SNIa,][]{Phillips1993}, Cepheids \citep{LeaPic1912}, Tip of the Red Giant Branch \citep[TRGB,][]{LeeFreMad1993}, and Surface Brightness Fluctuations \citep[SBF,][]{TonSch1988} among others. However, objects with distances measured with these more accurate methods typically make up only a small fraction of most peculiar velocity catalogs.  

Peculiar velocities of individual objects in a catalog can be combined to find the average velocity (bulk flow) of large volumes that contain some or all of the catalog objects.  The value of the bulk flow of a volume of some radius estimated from a catalog can then be compared to that predicted by the standard model of cosmology  \citep[$\Lambda$CDM][]{PlanckCosPar16}.  It is important to remember that theoretical models predict only the variance of the bulk flow components, since in a homogeneous and isotropic universe the average bulk flow should vanish.   This means that
adding new distances in order to measure the bulk flow on a given scale more accurately does not usually improve the constraint that the bulk flow can put on models unless the bulk flow is either increased or already larger than expected.  However, acquiring more data in order to make an accurate estimation of the bulk flow in a larger volume has the potential to strengthen constraints, since the $\Lambda$CDM model suggests that the variance for the bulk flow of a volume decreases with radius.   At a large enough radius, even a modest bulk flow can be in tension with the expected scale of the bulk flow.

In Section~\ref{sec:distvel} we discuss our unbiased distance and velocity estimators; in Section~\ref{sec:BFest} we present our methodology for estimating the bulk flow in a spherical volume using radial peculiar velocities; in Section~\ref{sec:data} we describe the \textit{CosmicFlows-4} (CF4) catalog that we are using in our analysis; in Section~\ref{sec:DisdInd} we discuss how we locate catalog objects in space; in Section~\ref{sec:analysis} we show how we estimate the bulk flow from the CF4 using the Minimum Variance (MV) method; in Section~\ref{sec:results} we present the results; we discuss our results in Section~\ref{sec:discussion}.

\section{Unbiased Distance and Velocity Estimators}
\label{sec:distvel}

Distances to galaxies are measured via distance moduli, $\mu$, with Gaussian distributed measurement errors $\sigma_\mu$.  The relationship between measured distance modulus and actual distance is
\begin{equation}
\mu = 5\log_{10}(d) + 25 + \delta,
\label{eq:mu}
\end{equation}
where $d$ is the distance in Mpc and $\delta$ is a measurement error drawn from a Gaussian distribution of width $\sigma_\mu$.  
We can use Eqn~\ref{eq:mu} to express an estimated distance $d_{est}$, which includes the effect of measurement noise,  in terms of the true distance $d$ and the measurement noise $\delta$,
\begin{align}
d_{est} &= 10^{\frac{\mu}{5} - 5} =  10^{5\log_{10}(d)/5 }10^{\delta/5} \nonumber\\
&= d e^{\delta\ln(10)/5}=d e^{\delta\kappa}
\end{align}
where $\kappa\equiv\ln(10)/5$. Averaging over measurement errors $\delta$ we find the well-known result that $d_{est}$ is biased, in that
\begin{equation}
\langle d_{est}\rangle = d\langle e^{\delta\kappa}\rangle \ne d.
\label{eq:av1}
\end{equation}
Using biased distance estimates results in similarly biased peculiar velocities.  Several methods have been proposed to deal with this bias \citep[\eg][]{WatFel15a}.  Here we take the novel approach of calculating the bias exactly so that it can be corrected for.  

Assuming that $\delta$ is drawn from a Gaussian distribution of width $\sigma_\mu$,  the probability density $P(\delta)$ of the measurement error $\delta$ is given by
\begin{equation}
P(\delta)= \frac{1}{\sqrt{2\pi}\sigma_\mu} e^{-\delta^2/2\sigma_\mu^2}.
\end{equation}
We can thus write the average in Eqn.~\ref{eq:av1} as
\begin{align}
\langle d_{est}\rangle &= d\frac{1}{\sqrt{2\pi}\sigma_\mu} \int_{-\infty}^{\infty} e^{\delta\kappa} e^{-\delta^2/2\sigma_\mu^2}\ d\delta \nonumber\\ &= d \frac{1}{\sqrt{2\pi}\sigma_\mu}  \int_{-\infty}^{\infty}\exp\left({\delta\kappa -\delta^2/2\sigma_\mu^2}\right)\ d\delta.
\end{align}
The integral can easily be evaluated by completing the square, giving
\begin{equation}
\langle d_{est} \rangle = d \ \exp\left({(\kappa\sigma_\mu)^2 /2}\right).
\end{equation}
Thus we can correct for the bias by multiplying our distance estimates calculated from the distance modulus by a factor, so that the corrected distance estimate $d_c$ given by 
\begin{equation}
d_c = 10^{\frac{\mu}{5} - 5}\ \exp\left(-{(\kappa\sigma_\mu)^2 /2}\right),
\label{eq:dc}
\end{equation}

is unbiased, in that $\langle d_c\rangle = d$.

To find the uncertainty in the distance estimate we need to find $\langle d_c^2\rangle$.  This is given by  
\begin{align}
\langle d_c^2\rangle &= d^2 \frac{f^2}{\sqrt{2\pi}\sigma_\mu} 
\int_{-\infty}^{\infty} e^{2\delta\kappa} e^{-\delta^2/2\sigma_\mu^2}\ d\delta \nonumber\\ 
&= d^2 \frac{f^2}{\sqrt{2\pi}\sigma_\mu}  \int_{-\infty}^{\infty}\exp\left({2\delta\kappa -\delta^2/2\sigma_\mu^2}\right)\ d\delta,
\end{align}
where we have defined the factor $f$ to be
\begin{equation}
f\equiv \exp\left({-(\kappa\sigma_\mu)^2 /2}\right).
\end{equation}

Completing the square as we did before gives
\begin{equation}
\langle d_c^2 \rangle = d^2 f^2  f^{-4} = d^2 f^{-2},
\end{equation}
thus, the uncertainty in the distance estimate $d_c$ is given by 
\begin{equation}
\sigma_{c}^2 = \langle d_c^2 \rangle- \langle d_c \rangle^2 = d^2 (f^{-2} - 1).
\end{equation}
The argument of the exponential in the factor $f$ is much less than 1 for typical uncertainties $\sigma_\mu$, so we can use the Taylor series expansion $e^x \approx 1+x$, giving
\begin{equation}
f^{-2} = \exp\left({2(\kappa\sigma_\mu)^2 /2}\right)\approx 1+(\kappa\sigma_\mu)^2,
\end{equation}
so that
\begin{equation}
\sigma_{c} \approx \kappa\sigma_\mu d.
\label{eq:unc}
\end{equation}
Thus distance uncertainties grow approximately linearly with distance and can be expressed as a percentage of distance.  

The unbiased distance estimate $d_c$ can be used to calculate peculiar velocities using
\begin{equation}
v_{c} = cz_{mod} - H_o d_c;
\label{eq:vc}
\end{equation}
where $H_o$ is the Hubble constant, $c$ is the speed of light, and $z_{mod}$ is the redshift modified to account for the deviation from a linear Hubble's Law,
\begin{equation}
z_{mod} = z\left( 1+ \frac{1}{2}(1-q_o )z-\frac{1}{6}(1- q_o- 3q_o^2 + j_o)z^2\right)
\end{equation}
where $z$ is the measured redshift and $q_o$  and $j_o$ are the deceleration and jerk parameters.  The use of modified redshift is important only for the more distant objects in the CF4 catalog.  
Peculiar velocities calculated in this way are also unbiased, in that 
\begin{align}
\langle v_c \rangle &= \langle  cz_{mod} - H_o d_c\rangle\nonumber = cz_{mod} - H_o \langle d_c\rangle\\ & = cz_{mod}-H_o d = v.
\end{align}
However, the errors in peculiar velocities estimated in this way are not Gaussian distributed.  

An alternative approach to calculating unbiased peculiar velocities was given in \cite{WatFel15a}.  In terms of the distance modulus $\mu$, that estimator can be written as 
\begin{equation}
v_{f}=  cz_{mod}\ln(10)\left( \ln(cz_{mod}/H_o) - \frac{\mu - 25}{5}\right),
\label{eq:vf}
\end{equation}
This formula gives peculiar velocities with Gaussian distributed errors; however, it has the disadvantage of only being unbiased for galaxies where the true velocity $v$ is much less than the redshift $cz$.  The uncertainty in this estimator is the same as that given in Eqn.~\ref{eq:unc}.  



\section{Estimating the Bulk Flow for a Spherical Volume using Radial Peculiar Velocities}
\label{sec:BFest}

The bulk flow is  one of the most basic ways to characterize the peculiar velocity field in the local Universe.   The bulk flow is defined as the average velocity in a region, usually taken to be a spherical volume $V$ surrounding the Milky Way galaxy with radius $R$:
\begin{equation}
U_i= \frac{1}{V}\int_V v_i\ d^3 x,
\label{eq:bf}
\end{equation}
where $v_i$ for $i=1,2,3$ are the cartesian components of the full three dimensional velocity field and $V=\frac{4}{3}\pi R^3$ is the volume of the sphere.  If we imagine the peculiar velocity field as being the sum of waves of various wavelengths $\lambda$, then the bulk flow averages out waves with $\lambda \lesssim R$ and thus mostly reflect the amplitudes of waves with $\lambda > R$.  Since homogeneity and isotropy requires the power spectrum to go to zero as $\lambda$ goes to infinity, the bulk flow should decrease with increasing radius $R$.  Thus the bulk flow is a probe of the power spectrum on scales that are difficult to probe using redshift surveys, and the measurement of the bulk flow provides an important test of the standard cosmological model.   

One difficulty in measuring the bulk flow as defined in Eqn.~\ref{eq:bf} is that we can only measure the radial component of the velocity field.  However, if we make the assumption that the velocity field is curl-free ($\vec\nabla\times \vec v= 0$), as velocities generated by gravity must be, then the radial velocity carries the same information as the full three dimensional field, and the bulk flow can be written as a weighted average of the \textit{radial peculiar velocity}.  The following derivation is taken from \cite{Nusser14,Nusser16}.  We begin by writing the velocity field in terms of a scalar potential $\phi$, 
\begin{equation}
\vec v = -\vec\nabla \phi.
\end{equation}
In terms of $\phi$ the bulk flow becomes  
\begin{equation}
U_i = -\frac{1}{V}\int_V \nabla_i \phi\ d^3 x = -\frac{1}{V}\int_V \vec\nabla\cdot (\hat x_i\phi)\ d^3 x,
\end{equation}
where $\hat x_i$ are the cartesian unit vectors.  The divergence theorem can then be used to convert the volume integral to an integral over the surface $S$ of the region,
\begin{equation}
U_i =  -\frac{1}{V}\int_S \hat r\cdot \hat x_i\phi\ d\Omega =  -\frac{1}{V}\int_S \hat r_i \phi\ R^2\ d\Omega,
\label{eq:div}
\end{equation}
where $\hat r$ is the radial unit vector with $\hat r_i$ being its $i$th component.   This integral picks out the dipole contribution to $\phi$.  We can make this explicit by expanding $\phi$'s angular dependence in terms of real-valued spherical harmonics
\begin{equation}
\phi({\bf r}) = \phi_o(r) + \sum_i \phi_i (r) \hat r_i + \sum_{l>1,m} \phi_{l,m}(r) Y_{l,m}(\theta,\phi),
\label{eq:expand}
\end{equation}
where the second term is a sum over the $l=1$ real valued spherical harmonics.   Plugging this into Eqn.~\ref{eq:div} and using the orthogonality of the spherical harmonics we obtain a simple expression for the bulk flow in terms of the value of $\phi_i$ on the surface of the region,
\begin{equation}
U_i = - \phi_i (R)/R.
\label{eq:bfphi}
\end{equation}

\cite{Nusser14} showed that the bulk flow can also be written as an integral of the radial peculiar velocity, $s$, times $\hat r_i$ and a radial weight function $w(r)$, 
\begin{align}
\frac{1}{V}\int_V w(r) s\ \hat r_i\ d^3 x &= -\frac{1}{V}\int_V w(r) \frac{\partial}{\partial r}\phi\ \hat r_i\ r^2 drd\Omega \\
&= -\frac{1}{R^3} \int_0^R w(r) \frac{\partial}{\partial r}\phi\ \ r^2 dr,
\end{align}
where we have plugged in Eqn.~\ref{eq:expand} and used the orthogonality of the spherical harmonics to do the integral over angles \citep[see also][]{PeeWatFel18}.  We see that  if $w = R^2/r^{2}$, then we can use the fundamental theorem of calculus to show that 
\begin{equation}
\frac{1}{V}\int_V w(r) s\ \hat r_i\ d^3 x = -\phi(R)/R=U_i,
\label{eq:bfr}
\end{equation}
thus showing that the bulk flow can be expressed as a weighted average of the \textit{radial} velocity with weights proportional to $\hat r_i/r^2$.

\section{Data}
\label{sec:data}

\begin{figure}
\centering
\includegraphics[scale=0.4]{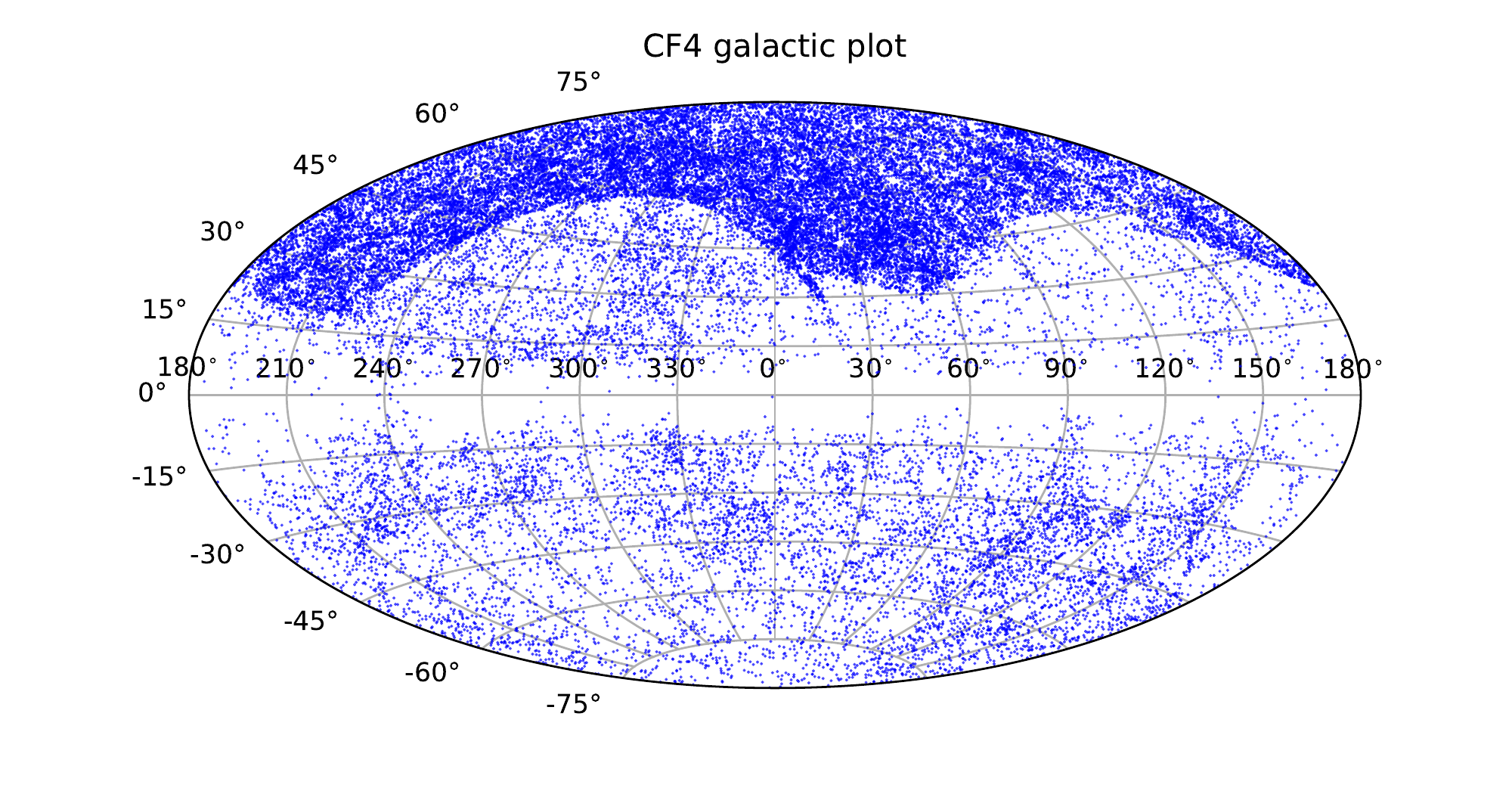}
\caption{Angular distribution of CF4 galaxies in Aitoff-Hammer projection Galactic coordinates.  Note the large number of galaxies in the northern Galactic hemisphere from the SDSS galaxy sample.}
\label{fig:CF4aitoff}
\end{figure}

In this paper we analyze the group version of the \textit{CosmicFlows-4} catalog \citep[][hereafter CF4]{CF4-full}.  The catalog gives modified redshift $cz_{mod}$, distance modulus $\mu$, uncertainty in distance modulus $\sigma_\mu$, Galactic longitude $l$, and Galactic latitude $b$, for over 38,000 groups and individual galaxies.  The majority of distance moduli in the CF4 are estimated using the TF \citep{TF77} and the FP \citep{dressler87} relations, with typical uncertainties of around 0.4 for individual galaxies.    The remaining distance moduli in CF4 are estimated using more accurate methods such as SNIa \citep{Phillips1993}, Cepheids \citep{LeaPic1912}, or Tip of the Red Giant Branch \citep{LeeFreMad1993}.  When there are multiple measured distance moduli for galaxies in a group, the group distance modulus is a weighted average and can have much smaller uncertainties.  The zero-point calibration of the CF4 catalog gives distances consistent with a value of $H_o=74.6$km/s/Mpc \citep{CF4-full}; however, it is important to note that our results are independent of the choice of zero-point of the catalog.

\begin{figure}
\centering
\includegraphics[scale=0.45]{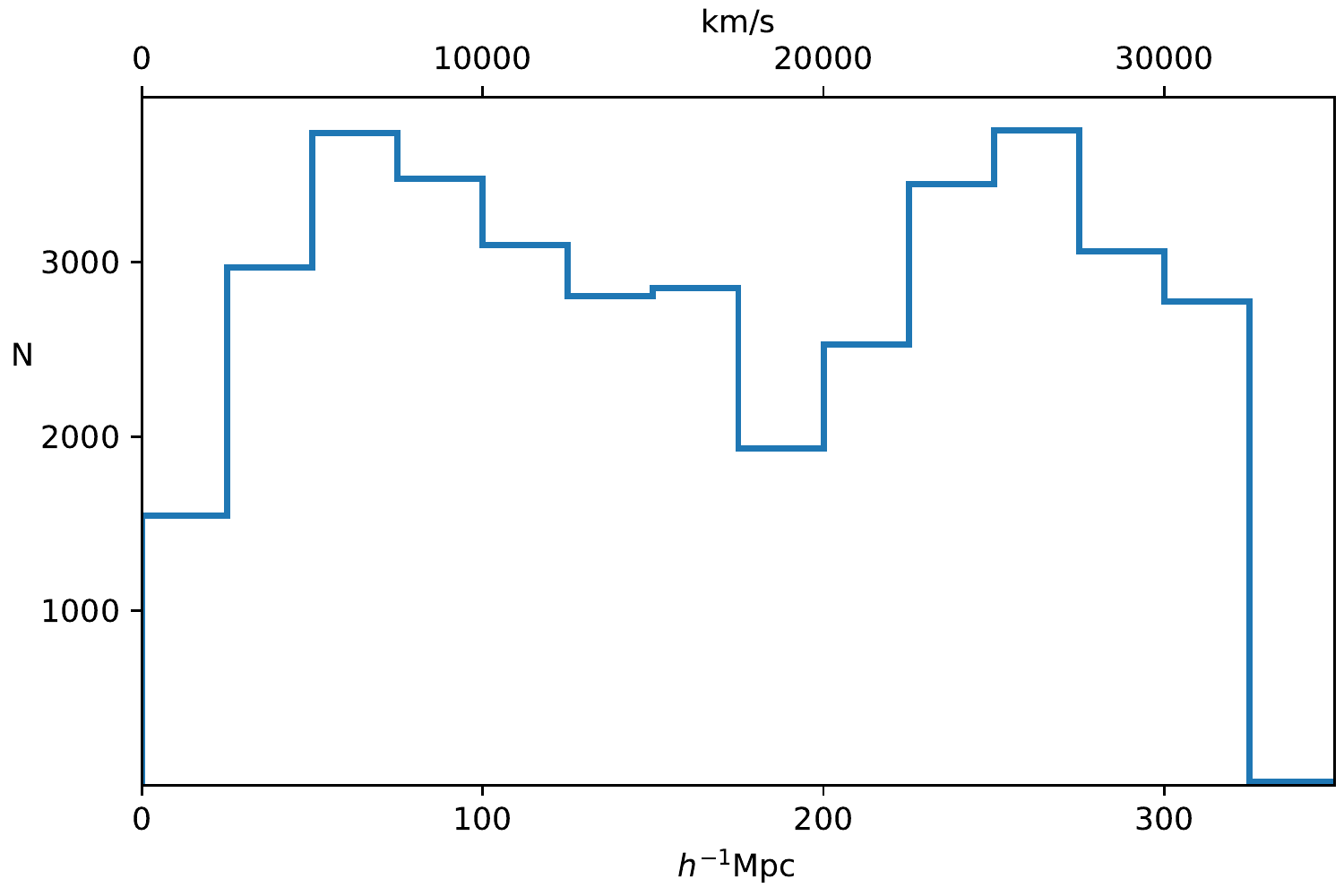}
\caption{The radial distribution of CF4 galaxies and groups.  }
\label{fig:rdist}
\end{figure}

\begin{figure}
\centering
\includegraphics[scale=0.45]{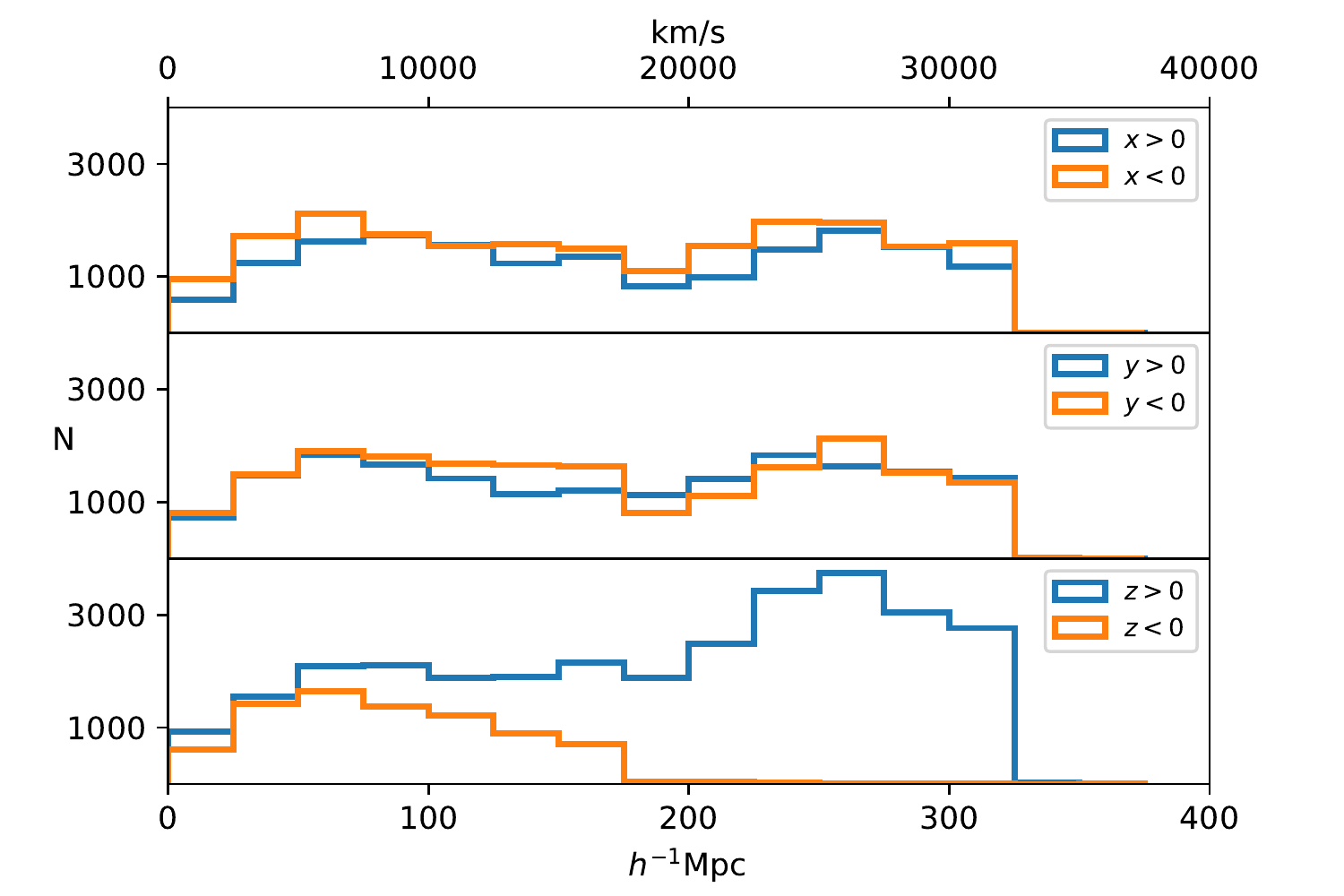}
\caption{The redshift distribution of CF4 galaxies in Galactic coordinates, dividing galaxies into those with positive and negative values of each cartesian coordinate.  We see that while galaxies are distributed fairly evenly in $x$ and $y$, in the $z$ direction the distribution is markedly different in the positive and negative directions.  This is mostly due to the SDSS galaxy sample. }
\label{fig:zdist}
\end{figure}

The CF4 adds two major new datasets to the \textit{CosmicFlows-3} \citep[CF3, ][]{CF3}: (a) distance moduli of $\simeq 34,000$ galaxies from the Sloan Digital Sky Survey \citep[SDSS, ][]{SDSS2000,HowlettSDSS} using the Fundamental Plane method, and (b) $\simeq 10,000$ distance moduli of spiral galaxies obtained using the Baryonic Tully Fisher Relation \citep{CF4btfr}.  The addition of the SDSS galaxies greatly expands the depth of the CF4 relative to the CF3; however, the SDSS galaxies are all in the northern Galactic hemisphere, so that the increase in depth is highly anisotropic.  In Fig.~\ref{fig:CF4aitoff} we show the angular distribution of the CF4 groups and galaxies in Galactic coordinates.  In Fig.~\ref{fig:rdist} we show the radial distribution of the CF4 objects.  In Fig.~\ref{fig:zdist} we show the radial distribution, but this time subdivided into objects in the positive and negative sides of each Galactic cartesian coordinate; this plot shows that while in the $x$ and $y$ directions groups and galaxies are fairly evenly distributed, in the $z$ direction the distribution of objects on the northern Galactic hemisphere is quite different than that in the southern hemisphere.  In particular, there are many more objects in the north and they are generally much deeper.   The anisotropic distribution of the CF4 galaxies makes it particularly important to use a formalism such as the minimum variance (MV) method that allows us to estimate a physically relevant bulk flow; otherwise the bulk flow would be difficult to interpret and not comparable to other results.  

\section{Locating Galaxies within the Volume}
\label{sec:DisdInd}

In order to theoretically model the data in the CF4 we need to first locate the groups in space.  While both galaxies and groups have accurate angular positions, distance estimates to galaxies are less accurate and require more thought.  For nearby objects,  Eqn.~\ref{eq:av1} provides an unbiased, accurate distance estimate.  However, as mentioned above, distance uncertainties grow linearly with distance, so that distances become increasingly uncertain as we consider more distant groups.  An alternative distance estimate $d_z$ is obtained from redshift information,  
\begin{equation}
d_z = cz_{mod}/H_o .
\label{eq:dz}
\end{equation}
Redshifts have small measurement uncertainties; the main source of error in $d_z$ is due to peculiar velocities, which contribute to the redshift through the Doppler effect.  For distant galaxies, the error in redshift distance due to peculiar velocities is smaller than the error in the distance obtained from the distance modulus.  

Here we use a novel way of determining the distance to galaxies to locate their positions in space.  We use a distance estimate that is a weighted average of the unbiased distance $d_c$ calculated from the distance modulus, given in Eqn.~\ref{eq:dc}, and the redshift distance $d_z$ given in Eqn.~\ref{eq:dz}, using weights that minimize the uncertainty in the average.  Our distance estimates are thus given by
\begin{equation}
d_{est} = (d_c/\sigma_c^2 + d_z/\sigma_v^2)/( 1/\sigma_c^2 + 1/\sigma_z^2),
\end{equation}
where the distance uncertainty $\sigma_c$ is given in Eqn.~\ref{eq:unc} and $\sigma_z = \sigma_v/H_o$, where $\sigma_v$ is the velocity dispersion of galaxies, which is about $300$km/s.   Practically speaking, $d_{est}$ corresponds to $d_c$ at small distances and to $d_z$ at large distances, making a smooth transition between the two for intermediate galaxies.  The exact value chosen for $\sigma_v$ does not have a significant effect on our results.  

To use our method, we also need to assign a peculiar velocity to each group.  For nearby objects, the estimate $v_c$, (Eqn.~\ref{eq:vc}), is preferred over $v_f$ (Eqn.~\ref{eq:vf}), since the formula used to calculate $v_f$ is not accurate at small distances.  However, at large distances the estimate $v_f$ might be preferred since it has Gaussian distributed errors and thus is a better match to the assumptions behind our method.   However, we experimented with using $v_c$ at small distances and then transitioning to $v_f$ at larger distances and it made virtually no difference to our results.  It appears that, at least for the current work, that as long as the velocities are unbiased, velocity errors are able average out effectively regardless of their distribution.   Thus in this work we use Eqn.~\ref{eq:vc} exclusively to estimate peculiar velocities.  

\section{Estimating the Bulk Flow using the Minimum Variance (MV) Method}
\label{sec:analysis}

Estimating the bulk flow is complicated by the fact that galaxies are both unevenly distributed in the volume and have varying uncertainties that generally increase with distance.  Simply calculating the weighted average of the radial velocities of galaxies within a given volume using weights proportional to $\hat r/r^2$ will not in general result in a good approximation to the integral over the velocity field given in Eqn.~\ref{eq:bfr}.  Instead, we imagine a more general weighted average
\begin{equation}
u_i = w_{i,n} s_n,
\label{bfest}
\end{equation}
where repeated indices are summed over, the $s_n$ are radial peculiar velocities of groups or individual galaxies in a survey, and $w_{i,n}$ are weights designed so that $u_i$ gives the best possible estimate of the bulk flow $U_i$ in a volume of radius $R$.  Here the weights $w_{i,n}$ should account for both the distribution of galaxies and the uncertainties of their peculiar velocity measurements.  To determine the optimal weights we will use the minimum variance (MV) method, developed in \cite{WatFelHud09,FelWatHud10,PeeWatFel18}.    The idea of the MV method is to calculate the weights that minimize the theoretical average square difference $\langle (u_i - U_i)^2\rangle$ between the estimated bulk flow components in Eqn.~\ref{bfest}, $u_i$, and the bulk flow components calculated for a theoretical ideal survey, $U_i$, consisting of exact peculiar velocity measurements measured at uniformly distributed points weighted by $1/r^2$.    In practice, this means generating a random set of $N_p$ points selected so that there are equal number of points per radial shell \citep[see][for details]{PeeWatFel18}.  

The method allows for constraints to be placed on the weights so that estimated bulk flow components are independent of the value of the Hubble constant.  This is particularly important given that we are in an era where the results of different methods of measurement of the Hubble constant are in tension with each other \citep[see \eg,][]{Freedman17,Di_Valentino2021,RieYuaMac22}.  

Once the weights for the bulk flow component estimates are determined, we can calculate the tensor angle-averaged window function ${\cal W}^2_{ij}(k)$ for the components, given by 
\begin{equation}
{\cal W}^2_{ij}(k) =w_{i,n}w_{j,m}f_{nm}(k),
\label{eq:wf}
\end{equation}
where repeated indices are summed over, and
\begin{equation}
 f_{nm}(k) = \int {d^2{\hat k}\over 4\pi} ( {\bf \hat r}_n\cdot {\bf \hat k} )( {\bf \hat r}_m\cdot {\bf \hat k} ) 
 e^{ ik\ {\bf \hat k}\cdot ({\bf r}_n - {\bf r}_m)}.
 \label{eq:fmn}
\end{equation}
The diagonal elements of the tensor window function ${\cal W}^2_{ii}(k)$ tell us which scales contribute to the bulk flow components.   In addition, a comparison of the window function for the bulk flow estimate $u_i$ with the window function for the ideal bulk flow $U_i$ can indicate how well our data can estimate the bulk flow of an ideal survey.   We discuss this comparison in more detail when we present our results in section \ref{sec:results}.  

Given the window functions in Eqn.~\ref{eq:wf} we can calculate the theoretical covariance matrix for the components of the bulk flow
\begin{equation}
R_{ij}=\langle u_i u_j\rangle ={H_0^2\Omega_{m}^{1.1}\over 2\pi^2}\int   dk\  P(k){\cal W}^2_{ij}(k),
\label{eq:Rij}
\end{equation}
where the window function ${\cal W}^2_{ij}(k)$ is given in Eqn.~\ref{eq:wf}.  Note that if distances are expressed in units of $h^{-1}$Mpc, the covariance matrix is independent of the value of $H_o$.  

The covariance matrix can be used together with the bulk flow component estimates $u_i$ to calculate a $\chi^2$ for the three component degrees of freedom,
\begin{equation}
\chi^2 = u_i R_{ij} u_j,
\end{equation}
where repeated indices are summed over.  From this $\chi^2$ distribution for three degrees of freedom we can find the probability of finding a value of $\chi^2$ that is as large or larger than our calculated value.

\section{Results}
\label{sec:results}

\begin{figure}
\centering
\includegraphics[scale=0.5]{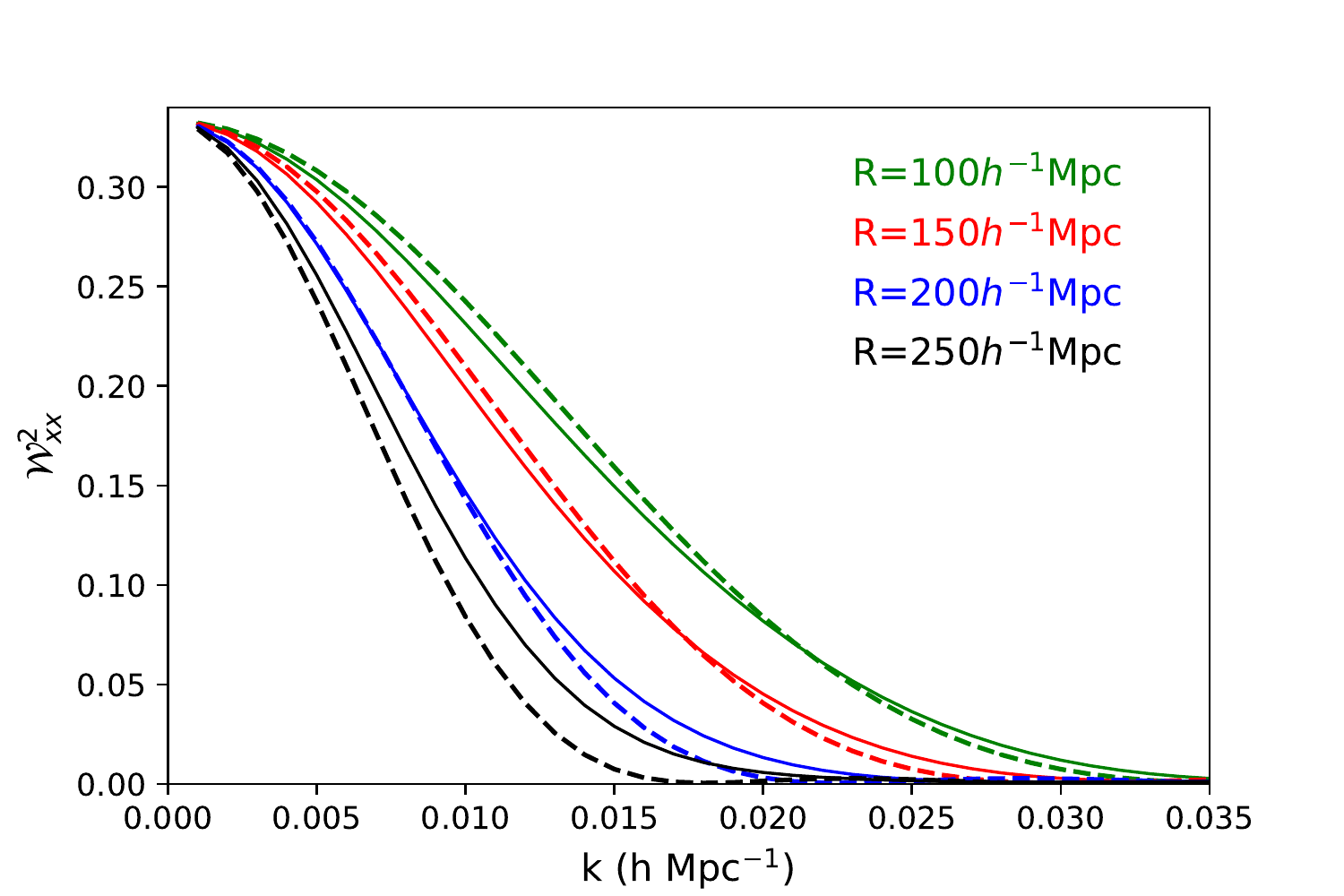}
\caption{Window functions for estimates of the $x$ component of the bulk flow for various radii $R$ calculated for the CF4 catalog.  The dashed lines show the window functions for an ideal survey of the same radius.}
\label{fig:wfx}
\end{figure}

\begin{figure}
\centering
\includegraphics[scale=0.5]{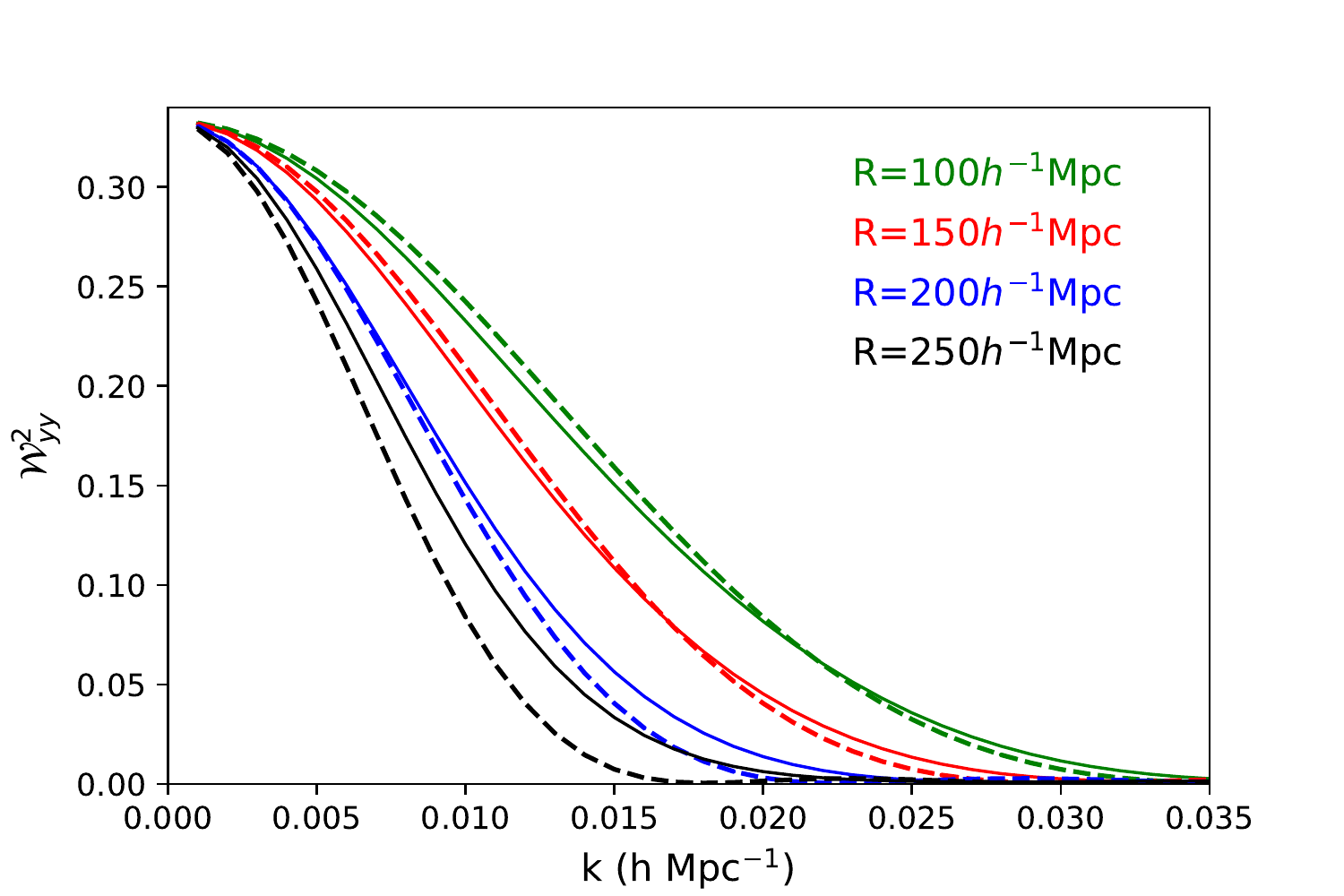} 
\caption{Same as Fig.~\ref{fig:wfx} for the $y$ component}
\label{fig:wfy}
\end{figure}
\begin{figure}
\centering
\includegraphics[scale=0.5]{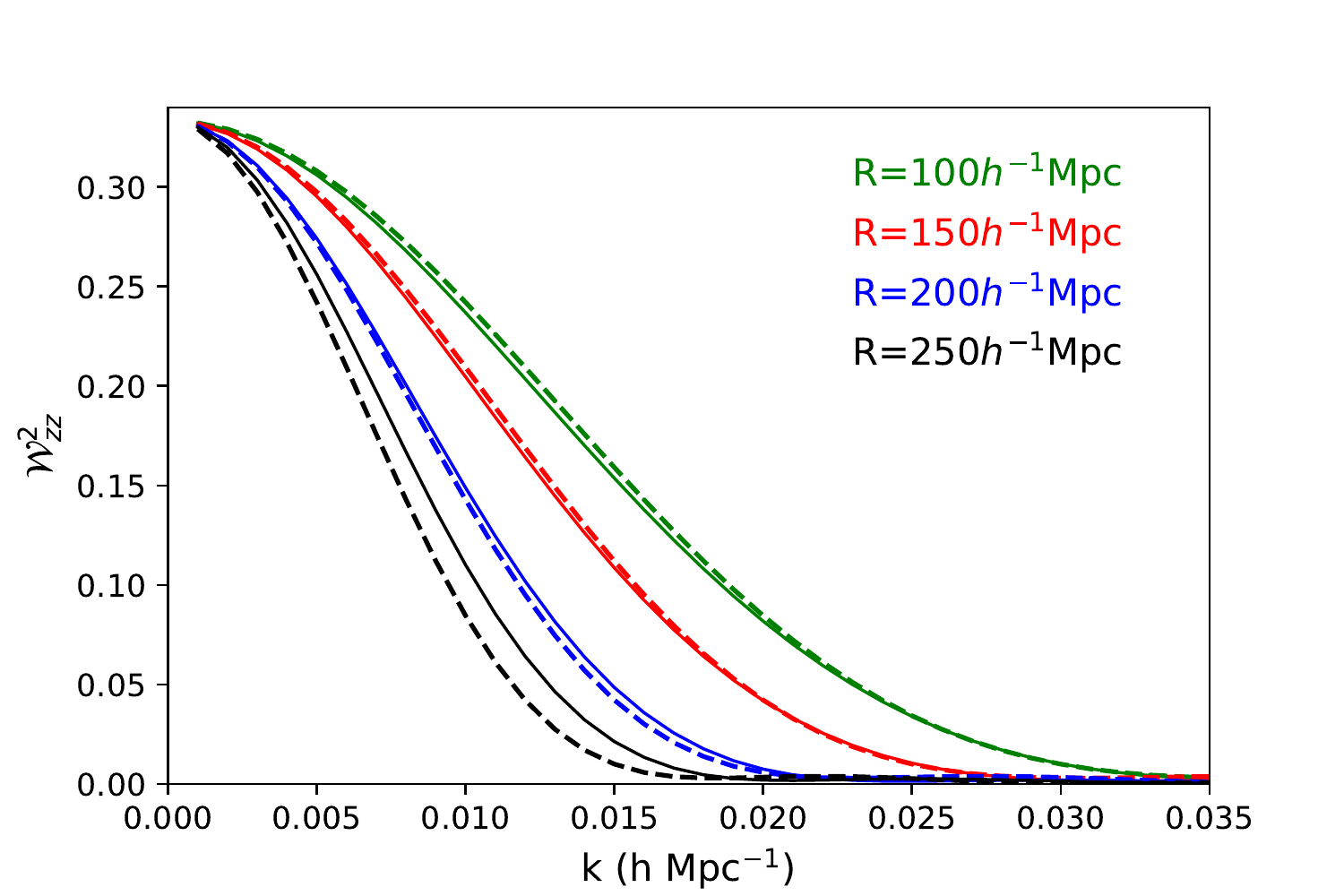}
\caption{Same as Fig.~\ref{fig:wfx} for the $z$ component}
\label{fig:wfz}
\end{figure}

As mentioned above, the CF4 is deeper than the CF3, allowing us to measure the bulk flow at greater depth, but also more anisotropic.  In order to determine the maximum depth to which we can measure the bulk flow accurately, we need to examine the window functions of our bulk flow estimates.  In Figs.~\ref{fig:wfx},~\ref{fig:wfy}, and ~\ref{fig:wfz} we show the window functions for the Galactic cartesian coordinates $x$, $y$, and $z$ bulk flow estimates for various radii $R$.  In the figures we see that the bulk flow estimate window functions match well with the ideal window functions out to a radius of about $R=200h^{-1}$Mpc, indicating that the estimates are probing the power spectrum in the same way.   Beyond this radius, we start to see a deviation (albeit small) of the estimated window function from the ideal window function, indicating that the catalog has insufficient information at the outer edge of the volume to accurately estimate the bulk flow on this scale.  In this work we focus on estimating the bulk flow for  $R=150h^{-1}$Mpc and $R=200h^{-1}$Mpc.  

\begin{figure}
\centering
\includegraphics[scale=0.55]{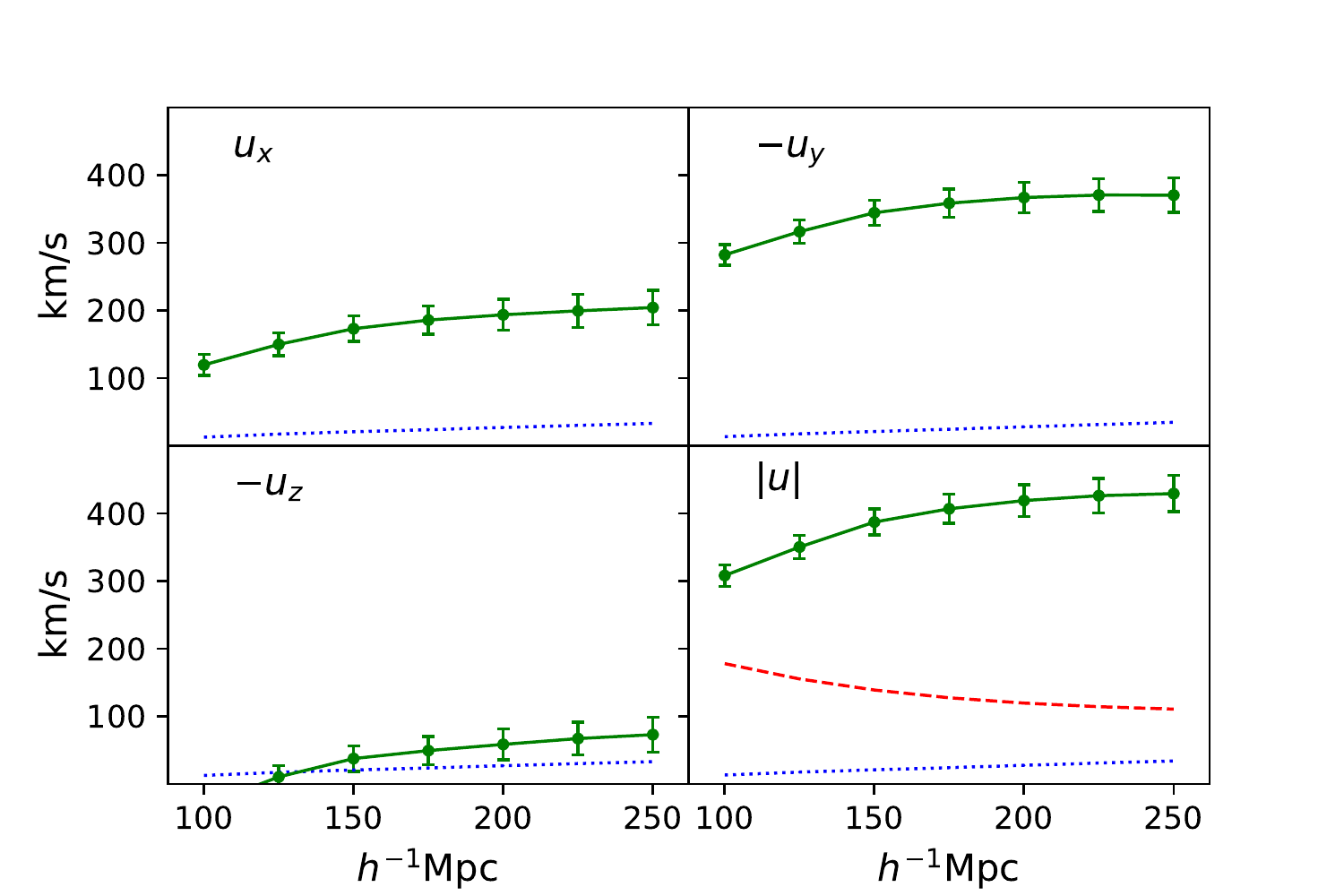}
\caption{The green points with error bars show the bulk flow components and magnitude estimated from the CF4 catalog as a function of radius $R$.  The error bars indicate the uncertainty in the estimates due to measurement noise. The dotted blue lines show the theoretical standard deviation of the expected differences between the bulk flow estimates and the bulk flow from an ideal survey calculated using the cosmological standard model (not including measurement noise).  The red dashed line indicates the theoretical expectation for the magnitude of the bulk flow calculated using the cosmological standard model.  }
\label{fig:bfR}
\end{figure}

In Fig.~\ref{fig:bfR} we show the bulk flow components and magnitude (with measurement noise error bars) calculated using the CF4 data as a function of the radius $R$.  Included in the plots (dotted blue lines) is the standard deviation of the expected difference between the bulk flow estimates and the bulk flow from an ideal survey calculated using the cosmological standard model with the \textit{Planck} central parameters \citep{PlanckCosPar16}; the total deviation from the ideal bulk flow is given by a quadrature sum of this value and the measurement noise.  The dashed red line on the magnitude plot indicates the expectation for the bulk flow magnitude calculated using the cosmological standard model.  

\begin{figure}
\centering
\includegraphics[scale=0.55]{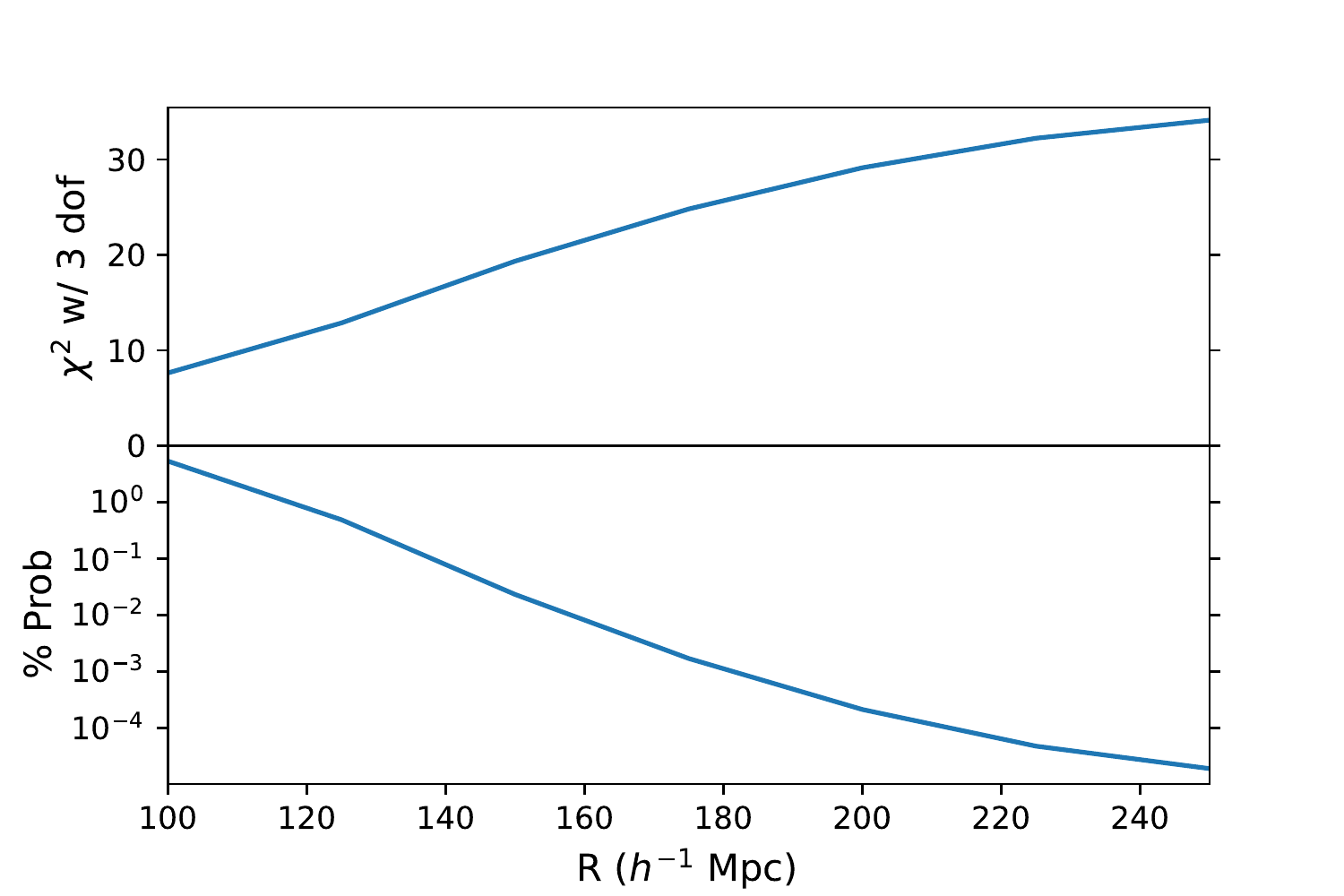}
\caption{ The $\chi^2$ for the three bulk flow degrees of freedom as a function of radius $R$.  Also shown in the corresponding probability to find a $\chi^2$ value that is as large or larger.}
\label{fig:chisq}
\end{figure}

In Fig.~\ref{fig:chisq} we show the $\chi^2$ for the three bulk flow component degrees of freedom as a function of $R$ calculated using the theoretical covariance matrix (Eqn.~\ref{eq:Rij}).  We also show the probability of finding a $\chi^2$ value that is as large or larger.  While we have focused on $R=150h^{-1}$Mpc and $R=200h^{-1}$Mpc in this paper, we can see from the figure that the probability of obtaining a bulk flow as large or larger continues to decrease as $R$ increases beyond $200h^{-1}$Mpc.  However, as discussed above, one must take our bulk flow results beyond $200h^{-1}$Mpc with a grain of salt, since at these radii the window functions for the bulk flow estimate are beginning to deviate from the ideal window functions.   This is mostly due to the lack of data at large distances; without much new information, bulk flow estimates on these scales will seem to change more slowly with radius than the actual bulk flow.  

\begin{table}
\caption{Summary of Bulk Flows for $R=150$\hmpc\ \ and $R=200$\hmpc.  The uncertainties include both the theoretical difference between the estimate and the bulk flow from an ideal survey and the measurement noise. \\ } 
\centering
{
\begin{tabular}{lccc}
\hline
& $R=150$\hmpc& $R=200$\hmpc \\
Expectation (km/s) & 139 & 120\\
Bulk Flow (km/s) & $387\pm 28$ & $419\pm 36$\\
Direction & $l=297^\circ\pm 4^\circ$ &$l=298^\circ\pm 5^\circ$\\
& $b=-6^\circ\pm 3^\circ$ & $b=-8^\circ\pm 4^\circ$ \\
$\chi^2$ with 3 d.o.f.  &19.34 & 29.13\\
Probability &0.023\% & 0.00021\%\\
\hline
\end{tabular}
}
\label{tab:prob}
\end{table}

\section{Discussion}
\label{sec:discussion}

As shown in Table ~\ref{tab:prob}, our bulk flow estimates in spheres of radii $R=150$\hmpc\ and $R=200$\hmpc\ have magnitudes in excess of $380$km/s and directions that are $\sim20^\degree-30^\degree$ away from the Shapley Concentration.  
We see that the probability of observing a bulk flow as large or larger for $R=150$\hmpc\ is small, only about 0.023\%.  This is significantly smaller than the probability of the CF3 bulk flow at this radius of 2.2\% \citep{PeeWatFel18};  the addition of new data has not only decreased the uncertainty but also increased the estimate of the bulk flow, resulting in a tension with the standard cosmological model at this radius that is significantly stronger.   Additionally, the fact that the CF4 is deeper than the CF3 allows us to accurately measure the bulk flow at larger radii.  As shown in Table~\ref{tab:prob}, the probability of obtaining a bulk flow as large or larger for $R=200$\hmpc\ is even smaller, 0.00021\%.  While this percentage does not quite reach $5\sigma$ significance, it does present a significant challenge to the standard model in addition to the tension seen in the value of the Hubble constant \citep[see \eg,][]{RieYuaMac22}.  

Most of the CF4 distances are obtained using either the Fundamental Plane (FP) or the Tully Fisher (TF) distance indicators.  In order to test whether these galaxies whose distances have been measured by different distance indicators give significantly different results, we have analyzed two smaller data sets: 1. where all groups and individual galaxies that contain only FP galaxies (26,800 groups and individuals out of 38,058) are removed; and 2. where groups and individual galaxies that contain only TF galaxies (9,060 groups and individuals out of 38,058) are removed.  Note that there are a few thousand groups containing galaxies with distance measurements from a mix of distance indicators that are common to both sub-catalogs.   The bulk flows estimated from both of these sub-catalogs do not differ significantly from that of the entire catalog (though the errors increase), giving us confidence that it is not a problem with one or the other of these main distance indicators that is causing the larger than expected bulk flow that we observe.  

Given the current tension in the value of the Hubble constant \citep[\eg,][]{Di_Valentino2021}, it is important to consider how a different value of $H_o$ might effect our results.  First, changing the value of $H_o$ introduces a spurious radial inflow or outflow into our data.  While a radial flow does not contribute to the bulk flow estimate in a completely isotropic survey, one might be concerned that in a survey such as the CF4 with different radial distributions in different directions, a radial flow may bleed into a bulk flow estimate.  However, as discussed in section~\ref{sec:analysis}, our method contains an explicit constraint making our bulk flow estimate independent of the value of $H_o$.  The second place that the value of $H_o$ might effect our results is by changing the scale of redshift distances $d_z$ used to locate groups in space, thus changing how the bulk flow probes the power spectrum.  Here, as is also discussed in section ~\ref{sec:analysis}, the power spectrum scales with $H_o$ in just such a way that if distances are measured in units of $h^{-1}$Mpc, theoretical bulk flow estimates are independent of the Hubble constant.  Taken together, these two considerations allow us to make bulk flow estimations, and more importantly, estimates of the probability of our bulk flows in the standard model of cosmology,  that are completely independent of the Hubble constant.  

Our results have clarified features of the bulk flow dependence on radius that were hinted at in previous studies \citep[\eg,][]{PeeWatFel18}.  In particular, we see from Fig.~\ref{fig:bfR} that, contrary to expectations, all three bulk flow components increase in magnitude as the radius of the volume increases beyond 100\hmpc.   This behavior is difficult to explain in the standard cosmological model, particularly since higher order moments of the velocity field do not appear to be larger than expected \citep{FelWatHud10}.  While this behavior is difficult to explain in the standard model, it could be an indication that the particle rest frame of the Universe is not the same as that inferred from the dipole in the cosmic microwave background (CMB) radiation \cite[see \eg,][]{KasAtrKoc08,MaGorFel11,KasAtr22}.  Indeed, significant evidence that the CMB frame is not the rest frame of the Universe can also be seen in the analyses of the distribution of Quasars on the sky \citep{SecVonRam21, SecVonRam22, DamLewBre22}.   We should note that the flattening out of the curves in Fig.~\ref{fig:bfR} is not necessarily due to a convergence to a rest frame; there is little information in the CF4 beyond 200\hmpc\ that could change the values of the bulk flow components.  

One might be concerned that our large bulk flow result is somehow arising from the anisotropic distribution of objects in the CF4 catalog.  There are two reasons to think that this is not the case.  First, the MV analysis method weights objects in such a way as to compensate for their distribution; regions with more objects are down-weighted and regions with fewer objects are up-weighted by the method.   The aim of the MV method is to generate weights to make the resulting bulk flow as close as possible to what would be obtained if there was a uniform distribution of objects in the volume.  We have checked that our method has not resulted in a small subset of objects having undue influence on our bulk flow result.   Second, Fig.~\ref{fig:zdist} shows that the anisotropy of the CF4 catalog is mainly in the $z$ direction;  the sample is much more balanced in the $x$ and $y$ directions.  Since we only have radial peculiar velocities, objects along a particular direction make the largest contribution to the bulk flow in that direction; this suggests that if anisotropy were effecting our result, it would have the greatest effect on the $z$-component of the bulk flow.  However, we see from Fig.~\ref{fig:bfR} that the $z$-component of the bulk flow is quite consistent with expectations; it is the $y$-component that is unexpectedly large in magnitude.    Altogether it is difficult to see how an anisotropy that is primarily in the $z$ direction could cause an unusually large bulk flow in the $y$ direction.

Given the greater context of our result being one of several that is contributing to an increasing tension with the standard model of cosmology, it is important to continue to improve our dataset and our analysis methods.  Unlike in the case of the Hubble constant, where resolving questions around the zero point of the distance ladder is essential, our result is relatively insensitive to the precise value of the Hubble constant and hence also the zero point.   Thus the accuracy of our result is likely to be improved only through the collection of more distance measurements.  Given that velocity errors increase with distance, finding distance estimators that give smaller percentage errors would greatly increase the value of new measurements.

\\

\noindent{\bf Acknowledgements:} 
RW, TA,CB, AR, and AW were partially supported by NSF grant AST-1907365.  HAF, RC, and YA-S were partially supported by NSF grant AST-1907404.  Funding for the \textit{Cosmicflows} project has been provided by the US National Science Foundation grant AST09-08846, the National Aeronautics and Space Administration grant NNX12AE70G, and multiple awards to support observations with HST through the Space Telescope Science Institute.

\bibliographystyle{mn2e}
\bibliography{haf}

\end{document}